\documentclass[%
 aip,
 jmp,%
 amsmath,amssymb,
 reprint,%
]{revtex4-1}

\usepackage{graphicx}
\usepackage{dcolumn}
\usepackage{bm}

\begin{document}

\preprint{AIP/123-QED}

\title{On Self Sustained Photonic Globes \footnote{Error!}}
\thanks{Author thanks the management of SNIST}

\author{K. Eswaran}
 \altaffiliation{SNIST, Jawaharlal Univ. of Tech., Yamnampet, Ghatkesar, Hyderabad 501301, India,\\ Formerly at Dept. of Theoretical Physics, Univ of Madras}
\date{\today}

\begin{abstract}
In this paper we consider a classical treatment of a very dense collection
of photons forming a self-sustained globe under its own gravitational
influence. We call this a {}``photonic globe'' We show that such
a dense photonic globe will have a radius closely corresponding to
the Schwarzschild radius. Thus lending substance to the conjuncture
that the region within the Schwarzschild radius of a black hole contains
only pure radiation.
As an application example, we consider the case of a very large photonic globe whose radius corresponds to the radius of the universe and containing radiation of the frequency of the microwave background (160.2 GHZ). It so turns out that such a photonic globe has an average density which closely corresponds to the observed average density of our universe.
\end{abstract}

\maketitle

\section{Introduction: Problem statement}
The possible existence of a self-sustained radiation existing as a spherical or near spherical region was first conjectured by J.A. Wheeler [1], who gave it the name Geon. Subsequently many researchers [2-6] have investigated this possibility and have studied two types of Geons - gravitational and electromagnetic. All the studies made were to find out if such structures can exist and be consistent with the field equations of general relativity. The conclusion arrived at was that such structures are essentially unstable and at best are not of long duration. In addition, Teo [7] has investigated the possibility of photons forming a stable orbit under the influence of a black hole and has concluded that such stable orbits are possible at a fixed distance which is exactly equal to 1.5 times the Schwarzschild radius of the black hole. Such structures were then called "photonic spheres", which actually is a thin layer of photons (like a thin ballon) at 1.5 times the distance of the Schwarzschild radius(SR). 

We in this paper  consider the possibility of photons existing under its own gravitational
field, and investigate under what conditions such a collection of
photons occupying a finite region including the origin,(photonic globe) can exist and be stable.

 In order to do so we make the
following assumptions:

(i) The treatment of the problem is classical

(ii) The velocity of each photon is the speed of light: \textit{c}

(iii) We assume that each photon is gravitationally attracted by another
photon, according to Newton's Gravitational law and behaves for this
purpose as having a {}``mass'' proportional to $h\nu/c^{2}$ ( $h$
and $\nu$ being Planck's constant, frequency of the photon resp.)

(iv) In this brief study it will be assumed that the photons all have
the same frequency, and that the photons form a self sustained globe
of radius R, the \textit{number density} of the photons $\sigma_{\nu}(r)$,
will be assumed to be a function of r alone, r being the radial distance
from the centre 0. 

It will be shown that by imposing conditions of stability of the system
one can show that the radius R of the photonic globe corresponds
to the Schwarzchilde radius, an expression for the number density
$\sigma_{\nu}(r)$ is also obtained.

\section{Brief details of calculation}

We herewith assume that a photonic globe, of radius R, consisting
solely of photons under its own gravitational field and centered about
the origin is extant. We define $M(r)$ to be the {}``mass'' of
an imaginary sphere of radius $r$, 0 $<$ r $\leq R$, then 

\begin{equation}
\ensuremath{M(r)=\int_{0}^{r}4\pi r^{2}\sigma_{\nu}(r)(h\nu/c^{2})dr}
\end{equation}

Now consider a point P at a distance r from the centre, O, and surrounded
by a volume element (in polar coordinates), the mass $\ensuremath{m_{\Delta}=\sigma_{\nu}(r)(h\nu/c^{2})r^{2}\, dr\, sin\theta\, d\theta\, d\phi}$

The gravitational force, F, on this small element $m_\Delta$  is
given by:

\begin{equation}
F=\frac{Gm_{\Delta}M(r)}{r^{2}}
\end{equation}

Now imagine the photons at P are moving inward with a velocity c and
making an acute angle $\psi$, with respect to the radial line drawn
from P to the origin O. Then the tangential velocity, $v_{t}$,
of the photons in this volume element will be $v_t = c \, sin \psi $.

The centrifugal force (c.f.) on this volume element will be given
by $ c.f=m_{\Delta}v_{t}^{2}/r $ , but $v_t^2 = c^2 \, sin^2 \psi $;
substituting the average value of $sin^2 \psi $ over 0 to $\pi$
as $1/2$, we see that $v_t^2 = c^2/2$, hence the cenrifugal force
on the photons in the volume element will be :

\begin{equation}
c.f.=\frac{m_{\Delta}c^{2}}{2\, r}
\end{equation}

The condition of stability requires that$F=c.f$ , hence by equating
(2) and (3), we have:

\begin{equation}
M(r)=\frac{c^{2}}{2G}\; r
\end{equation}

Substituting for $M(r)$, from (1), we have

\begin{equation}
\int_{0}^{r}r^{2}\sigma_{\nu}(r)\, dr\;=\left(\frac{c^{4}}{8\pi Gh\nu}\right)\: r
\end{equation}

Since eq. (5) must be true for all r, we can see that this is not
possible unless the number density, $\sigma_{\nu}(r)$, is given by
the following expression:

\begin{equation}
\sigma_{\nu}(r)\:=\left(\frac{c^{4}}{8\pi Gh\nu}\right)\:\frac{1}{r^{2}}
\end{equation}

It may be noted that though the number density seems to become infinite
as r tends to zero, the number of photons in a very small sphere of
radius $\epsilon$ will be $\sigma_{\nu}(\epsilon)\,\frac{4\pi}{3}\,\epsilon^{3}$
which is finite.

Now if we substitute $r=R$, and noting that $M(R)=M$, the mass of
the photonic globe, we have

\begin{equation}
R=\frac{2GM}{c^{2}}
\end{equation}

It may be noted that the rhs of (7) is nothing but the Scwarzchilde
radius.

\section{On the properties of the photonic globe}

From the above calculation, it so turns out that the radius R of a
photonic globe, eq.(7), is nothing but the Schwarzschild radius an
expression for which radius was derived by Schwarzschild in 1916,
for a spherically symmetric body by using equations of general realtivity
for regions \textit{outside} this radius. We have derived the same
expression for the Schwarzchilde radius by using completely different
arguments for regions \textit{inside} this radius by considering a collection of photons and using some assumptions detailed
above. It is well known that the event horizon for a black hole occurs
at a radius equal to the Scwarzchilde radius. The physics within this
radius is not well known and can only be guessed at. Also when eq(4)
written as $r=2GM(r)/c^{2}$, is a valid equation for any radius r
centered around the origin, $M(r)$ being the mass of the imaginary
sphere of this radius, we see that r is the Schwarzschild radius for
this sphere, so every point P at an arbitrary distance r in the globe
lies on an {}``event horizon''. The photon number density $\sigma_{\nu}(r)$,
as a function of r, of such a photonic globe is given by eq.(6).

The above calculation seems to lead to an interesting conjecture:
That the region within the Schwarzschild radius of a black hole consists
of pure radiation, a stable photonic globe, sustained within itself
by its own {}``gravitational'' field.

\section{Application regarding the density of the universe}

In this section, we will consider a very large photonic globe which contains photons corresponding to 160.2 GHZ, the frequency of the background radiation and assume the radius of the globe to be the radius of the universe.
If we start from the expression, Eq(6), for the number density of photons
at frequency $\nu$, and substitute $\upsilon\equiv\nu_{B}=160.2$GHz , where $\nu_{B}$is
the frequency of the background radiation of the universe (which corresponds
to a wave length $\lambda=0.1872$ cms), using the value of the
Gravitational constant $G=6.673\,\,10^{-8}$ cgs(cm-gram-second )
units and Planck's constant $h=6.626\,\,10^{-27}$erg-sec (cgs units
) , and the value of the  velocity of light $c=3.0\,\,10^{10}$ cms per sec; we see 
that at this frequency $\nu_{B}$ we can write 

\begin{equation}
\sigma_{\nu_{B}}(r)=4.55\,.\,10^{62}\,\,\frac{1}{r^{2}}
\end{equation}

which we denote for convenience as $\sigma_{\nu_{B}}(r)=\beta/r^{2}$,
where $\beta=4.55\,.\,10^{62}$ .

Now to calculate the total number of Photons $N_{R}$ inside a sphere
of radius R, we need to integrate the above and obtain 

\[
N_{R}=\int_{0}^{R}4\pi r^{2}\,\,\sigma_{\nu_{B}}(r)\, dr
\]
\begin{equation}
=4\pi\,\beta\, R
\end{equation}

The total energy, $E_{total}$, of radiation is then $E_{total}=N_{R}.h\upsilon_{B}$,.
The equivalent mass will be $M=E_{total}/c^{2}$. Hence the average
{}``mass density'' inside this sphere will be $\rho_{av}=M/(\frac{4\pi}{3}R^{3})$. That is

\begin{equation}
\rho_{av}=\frac{3\beta h\nu_{B}}{R^{2}c^{2}}
\end{equation}

\textbf{Now if we take R as the the radius of the visible universe
R=13.5 billion light years, ie. $R=1.227\,10^{28}$cms.}

Substituting these valuses for R, $\beta$ , $\nu_B$, and c, we get
an average mass density for the universe as $\rho_{av}=9.869\,\,10^{-30}$grams/cc.
Which is very close to the actual estimated mass density, by the WMAP.[8], as may be gathered
from the following quotation in the NASA, article [8]:

\textit{{}``WMAP determined that the universe is flat, from which
it follows that the mean energy density in the universe is equal to
the critical density (within a 0.5\% margin of error). This is equivalent
to a mass density of $9.9  \,\,  10^{-30} g/cm^3$, which is equivalent
to only 5.9 protons per cubic meter.''}

It is well known that only about 4 percent of the mass of the universe consists of Baryonic matter. So, if we, (for a stating approximation), adopt the hypothesis that the universe is a photonic globe containing photons of frequency of the CMB, then the above calculations give the correct average density, and obviously the correct Mass for the universe. However, if one considers the number density of photons of frequency $\nu_B $, it turns out to be grossly over estimated, (as can be easily calculated from the above equations) from the actual value of 400 photons per cc (near earth). 
So here we have a situation where the mass and mass-density are correct but the number of photons estimated are far too much. So where have all the extra photons gone?  It could then be conjectured that some unknown physical process has converted all this extra radiation into dark matter and  dark energy, thus keeping the total energy (mass) unchanged. Only further experimentation and research can resolve such issues.

\section{conclusion}
In this paper, we have considered the possibility of the existence of a stable a selfsustained
photonic globe and have arrived at the following: (i) that such a globe must
have its radius equal to the Schwarzchild radius and (ii) if we consider a photonic globe
which contains photons of frequency equal to 160.2 GHZ and a radius equal to the radius
of the universe then the average mass (energy) density of such a photonic globe is very
close to the latest estimate of the average mass density of the universe by NASA's WMAP
team[7].

\section{References}
1.Wheeler, J. A. (1957). ``Geons'' Physical Review 97 (2): 511. Bibcode 1955 PhRv...97..511W. doi:10.1103/PhysRev.97.511.

2.Brill, D. R.; Hartle, J. B.(1964).``Method of the Self-Consistent Field in General Relativity and its Application to the Gravitational Geon'' Physical Review 135 (1B): B271.Bibcode 1964 PhRv..135..271B. doi:10.1103/PhysRev.135.B271.

3.Louko, Jorma; Mann, Robert B.; Marolf,Donald(2005).``Geons with spin and charge'', Classical and Quantum Gravity 22 (7): 1451-1468. arXiv:gr-qc/0412012. Bibcode 2005CQGra..22.1451L.doi:10.1088/0264-9381/22/7/016. 

4.Perry, G. P.; Cooperstock, F. I. (1999). ``Stability of Gravitational and Electromagnetic Geons'' Classical and Quantum Gravity 16 (6): 1889 -916. arXiv:gr-qc/9810045. Bibcode 1999CQGra..16.1889P. doi:10.1088/0264-9381/16/6/321.

5.Anderson, Paul R.; Brill, Dieter R. (1997). ``Gravitational Geons Revisited'' Physical Review D 56 (8): 4824-4833. arXiv:gr-qc/9610074. Bibcode 1997PhRvD..56.4824A. doi:10.1103/PhysRevD.56.4824.

6.Teo, Edward (2003). "Spherical Photon Orbits Around a Kerr Black Hole". General Relativity and Gravitation 35 (11): 1909–1926. Bibcode 2003GReGr..35.1909T. doi:10.1023/A:1026286607562. ISSN 0001-7701.

7.WMAP Science Team, "Cosmology: The Study of the Universe," NASA's Wilkinson Microwave Anisotropy Probe,2013,
http:map.gsfc.nasa.gov/universe/WMAP\_ Universe.pdf, or http:map.gsfc.nasa.gov/universe/


\bibliography{ke-ph}

\end{document}